\documentclass [11pt]{article}
\usepackage{amsmath,amsthm,amsfonts,amscd,eucal,latexsym,amssymb}
\usepackage{epsfig}  
\def\loc{L^1_{loc}(\Sigma,d\Sigma)}

\def\cA{{\ca A}}

\def\cD{{\ca D}}

\def\cO{{\ca O}}

\def\cS{{\ca S}}

\def\cT{{\ca T}}
\def\cW{{\ca W}}

\def\bC{{\mathbb C}}           
 
\def\bI{{\mathbb I}}

\def\bR{{\mathbb R}}

\def\gA{{\mathfrak A}}       

\def\beq{\begin{eqnarray}}
\def\eeq{\end{eqnarray}}
\def\pa{\partial}
\def\at{\left(}               
\def\aq{\left[}               
\def\ag{\left\{}              

\def\ct{\right)}              
\def\cq{\right]}              
\def\cg{\right\}}             
\newcommand{\ca}[1]{{\cal #1}}         
\def\bg{{\bf g}}             

\def\ga{\gamma}
\def\de{\delta}

\def\la{\lambda}

\def\si{\sigma}
\def\om{\omega}

\def\De{\Delta}

\def\Si{\Sigma}
\def\Om{\Omega}

\def\lloc{{\sf Loc}}
\def\loc{{\sf CLoc}}
\def\alg{{\sf Alg}}
\def\testf{{\sf Test}}
\def\talg{{\sf TAlg}}
\def\WF{{\text{WF}}}

\newcommand{\nref}[1]{(\ref{#1})}

\def\supp{supp\;}

\newcounter{proposition}[section]
\newcounter{theorem}[section]
\newcounter{lemma}[section]
\newcounter{definition}[section]
\newcounter{corollary}[section]
\def\theproposition{\thesection.\arabic{proposition}}
\def\thetheorem{\thesection.\arabic{theorem}}
\def\thelemma{\thesection.\arabic{lemma}}
\def\thedefinition{\thesection.\arabic{definition}}
\def\thecorollary{\thesection.\arabic{corollary}}

\newcommand{\se}[1]{\section{#1}}

\def\vsp{\vspace{0.2cm}}
\def\vspp{\vspace{0.1cm}}

\def\sse #1 {\vsp\ifhmode{\par}\fi\refstepcounter{subsection}
  \noindent {\bf\thesubsection}. {\em #1}.\quad
  \addcontentsline{toc}{subsection}{\protect\numberline{\thesubsection} #1}%
  }

\def\ssb #1 {\vsp\ifhmode{\par}\fi\refstepcounter{subsection}
  \noindent {\bf\thesubsection.} {\bf #1.}\quad
  \addcontentsline{toc}{subsection}{\protect\numberline{\thesubsection} #1}%
  }

\def\ssa #1 {\ifhmode{\par}\fi\refstepcounter{subsection}
  \noindent {\bf\thesubsection.} {\bf #1.}\quad
  \addcontentsline{toc}{subsection}{\protect\numberline{\thesubsection} #1}%
  }

\def\proposizione #1 {\vsp\ifhmode{\par}\fi\refstepcounter{proposition}
  \vsp\ifhmode{\par}\fi\noindent {\bf Proposition \theproposition}. \quad {\em #1}}
\def\teorema #1 {\vsp\ifhmode{\par}\fi\refstepcounter{theorem}
  \vsp\ifhmode{\par}\fi\noindent {\bf Theorem \thetheorem}. \quad {\em #1}}
\def\lemma #1 {\vsp\ifhmode{\par}\fi\refstepcounter{lemma}
  \vsp\ifhmode{\par}\fi\noindent {\bf Lemma \thelemma}. \quad {\em #1}}
\def\definizione #1 {\ifhmode{\par}\fi\refstepcounter{definition}
  \vsp\ifhmode{\par}\fi\noindent {\bf Definition \thedefinition}. \quad {\em #1}}
\def\corollario #1 {\vsp\ifhmode{\par}\fi\refstepcounter{corollary}
  \vsp\ifhmode{\par}\fi\noindent {\bf Corollary \thecorollary}. \quad {\em #1}}

\def\proof #1 {\vspp\ifhmode{\par}\fi\noindent {\it Proof.} {#1} $\Box$\vsp\par}
\def\remark {\vsp\ifhmode{\par}\fi\noindent\noindent {\bf Remark:} 
}

\def\Mi{M_1}

\def\Mii{M_2}

\begin{document}

\hfill{\sl Desy 08-070, ZMP-HH/08-10, June 2008} 
\par 
\bigskip 
\par 
\rm 
 
\par 
\bigskip 
\LARGE 
\noindent 
{\bf Conformal generally covariant quantum field theory: The scalar field and its Wick products.} 
\bigskip 
\par 
\rm 
\normalsize 

\large
\noindent{\bf Nicola Pinamonti$^{a}$}\\
\par

\small
\noindent
II. Institut f\"ur Theoretische Physik, Universit\"at Hamburg,
Luruper Chaussee 149, 
D-22761 Hamburg, Germany.\smallskip

\bigskip

\noindent $^a$  nicola.pinamonti@desy.de \\ 
 \normalsize

\par 
 

\rm\normalsize 
 
 
\par 
\bigskip 

\noindent 
\small 
{\bf Abstract}.
In this paper we generalize the construction of generally covariant quantum theories given in \cite{BFV} to encompass the conformal covariant case.
After introducing the abstract framework, we discuss the massless conformally coupled Klein Gordon field theory,
showing that its quantization corresponds to a functor between two certain categories.
At the abstract level,  the ordinary fields, could be thought as natural transformations in the sense of category theory.
We show that, the Wick monomials without derivatives (Wick powers), can be interpreted as fields in this generalized sense, provided a non trivial choice of the renormalization constants is given.
A careful analysis shows that the transformation law of Wick powers is characterized by a weight, and 
it turns out that the sum of  
fields with different weights breaks the conformal covariance.
At this point there is a difference between the previously given picture due to the presence of a bigger group of covariance.
It is furthermore shown that the construction does not depend upon the
scale $\mu$ appearing in the Hadamard parametrix, used to regularize the fields.
Finally, we briefly discuss some further examples of more involved fields.

\normalsize
\bigskip 

\se{Introduction}
The systematic  analysis of quantization in terms of functors given by Brunetti, Fredenhagen and Verch \cite{BFV}, opened an interesting new way to interpret the quantum field theory on curved spacetimes.
With this new ideas, the expectation values of fields in different spacetimes can be compared in a mathematically rigorous way.
Some interesting new applications have been developed following this line of thinking, we remind here the work of Buchholz and Schlemmer \cite{BS} and Schlemmer and Verch \cite{SV}, where the authors   deal  consistently with expectation values of fields in different spacetimes.
Another interesting use of similar ideas can be found in the derivation of local energy bounds in curved spacetime as performed by Fewster \cite{Fewster}. 
The use of these concepts plays a central role in the development of a perturbative theory of quantum gravity as well, to this end we would like to remind the interesting paper of Brunetti and Fredenhagen \cite{BF}.

A central role in the analysis performed in \cite{BFV} is played by the study of the isometric embeddings between different spacetimes and their interplay with the quantization procedure. It was shown that the quantization of the massive Klein Gordon fields can be encompassed in the new scheme. Furthermore, the field itself and its Wick powers, as constructed by Hollands and Wald in \cite{HW01,HW02,HW05}, can be interpreted as generally covariant quantum fields.
Here we would like to address the same problem in the case of field theories having a larger group of symmetry, namely the locally conformally covariant case.
Hence, we introduce the notion of  generally conformally covariant fields by enlarging the abstract setup presented in \cite{BFV}.
The idea of considering more complicated morphisms than isometries  appeared for the first time in the work of Brunetti \cite{Romeo},  we would like to follow similar line of reasoning.

If the extension of the covariance to the conformal covariance is expected to hold true at the level of canonical commutation relations and hence at the level of the simple scalar field, the situation is expected to be different considering the extended algebra of fields, namely the fields defined by means of a regularization.
It usually happens that the regularization breaks the conformal covariance,
technically speaking this is due to the unavoidable presence of a length scale in the Hadamard parametrix used to regularize the fields.
It is then an unexpected fact, that, in the four dimensional case, despite the presence of this length scale and more generally despite the presence of quantum anomalies, a proper but large subset of the algebra of local fields, contains locally conformally covariant fields.
%
%
%
%
%
%
We shall show that the Wick powers (the Wick monomial without derivatives) are contained in this subset, provided a non trivial choice of renormalization freedom is performed\footnote{A detailed analysis of the renormalization freedom can be found in the work of Hollands and Wald 
\cite{HW01,HW05}.}.
At this point it seems interesting to remark that the requirement of being conformally covariant restricts the renormalization freedom usually present in the construction of these fields.
This fact seems to be a peculiarity of the four dimensional case, it is in fact known that, for example, in the two dimensional case, the Wick powers ($\varphi^n$) are not locally conformally covariant (they are not primary in the language of CFT \cite{DMS}), we shall furthermore comment on this restriction in a subsection devoted to the analyses of the extension of this results to general dimensions. 
Another interesting difference that arises in the case under investigation is that the transformations rules enjoyed by the Wick powers are characterized by the presence of a weight. 
Furthermore, the sum of Wick monomials with different weight breaks the conformal covariance.

The analysis performed in this paper allows to 
geometrically relate  a larger class of spacetimes than in \cite{BFV}, namely those that are locally connected by a conformal transformation.
%
%
In this way it is possible, for example, to transplant observables (and states) from the de Sitter spacetime to the Minkowksi one. This could be useful in the study of concepts like local equilibrium states \cite{BOR} in the case of conformally covariant theories as well.

The paper is organized as follows: at first
we introduce the notion of locally generally conformal covariant quantum fields.
The example of the massless conformally coupled scalar Klein Gordon field is studied in the second section, we shall present the transformation rule of the fundamental solutions and of the Hadamard parametrix in particular.
The third section contains the analysis of the Wick powers in four dimensions and a subsection devoted to the discussion of the differences between this case and the case of spacetimes with general dimensions.
Some final comments and some further non trivial examples of more complicated fields are given in the fourth section. 
The appendix contains some technical computation used in the derivation of the results.

\ssb{Categorical formulation of locally conformally covariant field theory}
We are going to enumerate the relevant categories that will be used later for the formulation of a conformal quantum  field theory in terms of a functor between certain categories.
Before doing it, we introduce some small modifications to the locally covariant picture of quantum field theory presented for the first time in \cite{BFV}, in order to adapt the formalism to include the case of conformal invariant theories.
The key obervation is that 
conformal invariant field theory should be invariant under a reacher group of transformations, namely the local conformal transformations.
It is interesting to notice that such transformations share a lot of nice properties with isometries, the causal structure is preserved by such transformations in particular and this fact will play a central role later on.
For a better formalization of these concepts we would like to introduce the notion of conformal embedding.
\definizione{Consider two globally hyperbolic spacetime $(\Mi,\bg_1)$ and $(\Mii,\bg_2)$ then, a map $\psi:\Mi\to \Mii$ is called {\bf conformal embedding} if it is a diffeomorphism between $\Mi$ and $\psi(\Mi)$ and
the push forward $\psi_*$ acts on the metric $\bg_1$ in the following way:
$\psi_* \bg_1  = \Om^{-2}\; {\bg_2}_{|_{\psi(\Mi)}}$ where $\Om$ is a strictly positive smooth function on $\psi(\Mi)$, called {\bf conformal factor}.}

\vsp
\noindent
In the following we shall consider the case of a conformal embedding $\psi$  between two globally hyperbolic spacetimes $(\Mi,\bg_1)$ and $(\Mii,\bg_2)$ that preserves orientation and time orientation and such that the image $(\psi(\Mi),\bg_2|_{\psi(\Mi)})$ is also an open globally hyperbolic subset of $(\Mii,\bg_2)$.
We would like to remark that, under the given hypotheses, $\psi$ preserves the causal structures of the spacetime\footnote{See the Appendix D of \cite{WaldBook} for more details}, mapping for example causal curves to causal curves and so on and so forth. 

At this point it seems important to stress a difference between the conformal embeddings used in this paper and the 
conformal transformations that form the so called conformal group.
The main difference arises because 
we are not simply considering coordinate transformations but general mappings between different spacetimes.
For example, in the four dimensional Minkowski spacetime, the conformal transformations that can arise as coordinate transformations form a finite-dimensional group $SO(2,4)$, while much more freedom
is allowed by conformal embeddings.

The following action of weighted conformal transformations on test functions will
play 
a distinguished role in the definition of the weight of the field.
\definizione{\label{actiontest} Let $\psi$ be a conformal embedding between $(\Mi,\bg_1)$ and $(\Mii,\bg_2)$ with conformal factor $\Om_\psi$ then, the {\bf weighted action on test functions} $\psi^{(\lambda)}_*$ is the map from $C^\infty(\Mi)$ to $C^\infty(\psi(\Mi))$ such that,
$$
\psi_*^{(\lambda)}(f)(x):= \Om_\psi^{-\lambda} (x) (f \circ \psi^{-1}) (x) .
$$
Where $\la\in \bR$ is called the {\bf weight of the map}.}

The previously given definition deserves some comments regarding its domain of definition and its inversion. While it is clear that $\psi_*^{(\lambda)}$ can also be thought as acting on compactly supported smooth function
$\psi_*^{(\lambda)}:C_0^\infty(\Mi)\to C_0^\infty(\Mii)$, 
that is not true anymore considering smooth functions,
in fact  
 $\psi(\Mi)$ is in general a proper subset of $\Mii$ hence a smooth function $f$ that is not compactly supported on $\Mi$ is not 
 mapped to a smooth function in $C^\infty(\Mii)$. It is indeed impossible to extend uniquely $\psi_*^{(\lambda)}(f)$ on $\Mii$ outside $\psi(\Mi)$.
Despite  the presence of these domain problems we would like to notice that $\psi_*^{(\lambda)}$ is invertible either on $C_0^\infty(\psi(\Mi))$ or on $C^\infty(\psi(\Mi))$.
The particular conformal embedding $\psi:(M,\bg)\to (M,\bg')$ such that every $p\in M$ is mapped to $\psi(p)=p$, is called {\bf conformal transformation}. Moreover, if the conformal factor $\Om_\psi$ of a conformal transformation is a constant then it is called {\bf rigid conformal transformation} or {\bf rigid dilation}.

\vsp

\noindent
We enumerate here the categories used later on; these definitions are very similar to those given in \cite{BFV}.
For this reason we shall stress, case by case, the differences we have to implement in order to encompass also the conformal transformations in the framework.

\begin {itemize}
\item[\loc:]
This is the category that encompasses all the geometric structures of the theory. 
The object of $\loc$ are all the four dimensional oriented and time oriented globally hyperbolic spacetimes.
While the morphisms are all the conformal embeddings $\psi:(\Mi,\bg_1)\to (\Mii,\bg_2)$
with the following additional properties, that are the same as previously given:
$(i)$ $(\psi(\Mi),\bg_2 |_{\psi(\Mi)})$ is an open globally hyperbolic subset of $(\Mii,\bg_2)$ and $(ii)$ the morphisms preserve orientation and time orientation\footnote{The requirement of global hyperbolicity for $\psi(\Mi)$ is equivalent to the requirement of 
causal convexity of $\psi(\Mi)$ in $\Mii$.
In other words every causal curve with endpoints in $\psi(\Mi)$ has to lie inside $\psi(\Mi)$ too.}.
The composition of morphisms is defined as the composition map of conformal embeddings in the usual way.
The category $\loc$ is an extension of the category $\lloc$ given in \cite{BFV}, in the sense that in $\loc$ there is a larger class of  morphisms then in $\lloc$.
\item[\alg:] 
There is no need to modify the category of $\alg$ introduced in [BFV].
The object of $\alg$ are all the $C^*$-algebras built on a globally hyperbolic spacetime $(M,\bg)$,
possessing the unit element, while their morphisms 
are the injective $*-$homomorphisms that preserve the unit; once again the composition descends from the usual composition map of $*-$homomorphism. 
\item[\talg:] The definition of a $\talg$ follows easily the one of $\alg$; the difference is that the object of this category are taken to be only $*-$algebras with unit, instead of $C^*$-algebras. There is no modification between this and the previously given definitions.
\item[$\testf^\la$:] The objects of this category are the sets of compactly supported smooth functions $C_0^\infty(M)$ on the spacetimes $(M,\bg)$. The morphisms 
are the 
weighted transformation $\psi_*^{(\lambda)}:M\to M'$ with a fixed $\lambda$
and their action is like the one presented in definition   \ref{actiontest}.

\end{itemize}

\noindent
It seems interesting to notice that the categories $\alg$ and $\talg$ are defined in the same way as on \cite{BFV,Romeo}, in a certain sense the algebraic formulation of quantum field theory is already suitable to describe conformal transformations.
Furthermore the scaling transformations have already been considered as geometric morphisms in the work \cite{Romeo}.

\ssb{Quantum Conformal Field theory as a Functor and Conformal fields as Natural transformations}\label{functor}
We are now in place to define the {\bf locally covariant conformal quantum field} as a functor between the  two categories $\loc$ and $\alg$, such that the objects of $\loc$ are mapped to the objects of $\alg$ whereas the morphisms $\psi$ of $\loc$ are mapped into the morphisms $\alpha_\psi$ of $\alg$, in such a way that the following diagram commutes
\begin{equation*}
\begin{CD}
(M,\bg) @>\psi>> (M',\bg')\\
@V{\cA}VV     @VV{\cA}V\\
\cA(M,\bg)@>{\alpha_\psi}>> \cA(M',\bg')
\end{CD}
\end{equation*}
and the following composition property holds:
$$
\alpha_\psi\circ\alpha_{\psi'}=\alpha_{\psi\circ\psi'}\; ,\quad \alpha_{\bI_M} = \bI_{\cA(M)}\;.
$$
The same construction can be repeated substituting the category $\alg$ with $\talg$. 

Despite the meaningfulness of the previously given definition 
and the presence of examples of the given framework, 
it is not at all clear if observables with a certain physical meaning in a spacetime are mapped to observables with the same meaning, on the other spacetime.
In general this is indeed not the case and it is precisely because of this problem that the ordinary fields need to be introduced in an alternative way. 
In the picture we are going to introduce, they will assume the particular meaning of natural transformations between categories.

To this end it is useful to consider the set of weighted test functions $\cD^\la$ as a functor between $\loc$ and $\testf^\la$.
More precisely let's indicate by $\cD^\la(M,\bg)$ the category whose elements are the sets of compactly supported smooth functions $C^\infty_0(M)$, and the morphisms $\alpha^\la_\psi$ between these sets are defined by means of the weighted action on test functions as defined in \ref{actiontest}. Clearly $\cD$ can also be seen as a functor between the category of $\loc$ to $\testf$.
We are now ready to introduce the notion of {\bf conformal quantum field as a natural transformation} between two functors.
\definizione{%
\label{conformalfield}
A field $\Phi^\lambda_{(M,\bg)}$ of weight $\la$ is a linear transformation between the functor that realizes the test functions $\cD^{4-\la}:(M,\bg)\to \cD^{4-\la}(M,\bg)$ and the functor that realizes the topological algebras $\cA:(M,\bg)\to\cA(M,\bg)$ such that the following diagram commutes
\begin{equation*}
\begin{CD}
\cD^{4-\la}(M,\bg)       @>\Phi^\la_{(M,\bg)}>>      \cA(M,\bg)\\
@V{\psi^{(4-\la)}_{*}}VV                   @VV{\alpha_\psi^\la}V\\
\cD^{4-\la}(M',\bg')   @>{\Phi^\la_{(M',\bg')}}>> \cA(M',\bg')
\end{CD}
\end{equation*}
}
The preceding definition can be written more explicitly by means of the following conformal covariance property:
$$
\alpha^\la_\psi(\Phi^\la_{(M,\bg)})(f)=\Phi^\la_{(M',\bg')}(\psi^{(4-\lambda)}_*(f))\;,
$$
where $\psi_*^{(\lambda)}(f)$ is defined as a weighted transformation as given in the definition \ref{actiontest}.
We call $\lambda$ the weight of the field $\Phi^\lambda$.

\noindent
The difference between the weight in the test functions and the weight in the fields can be understood 
taking into account the transformation rule enjoyed by the 
volume form. Under a conformal embedding $\psi:(M,\bg)\to(M',\bg')$, 
$$
\sqrt{\bg'(\psi(x))}\Om^{-4}(\psi(x))=\sqrt{\bg(x)}
$$
where $\bg'$ stands for the determinant of the metric computed in a chart of $M'$ containing $\psi(x)$ and $\bg$ is for the determinant of $\psi_* \bg $ computed in the same chart.
%

As a consequence of the given definitions, 
linear combinations of fields with different weights are not 
conformally covariant fields.
Precisely at this point there is a great difference with what was addressed in \cite{BFV}, where also the linear combinations of fields with different ``weights'' were taken into account.

In two dimensional conformal field theory the fields that posses this property are called {\bf primary fields} \cite{DMS}.
Hence there is a relation between the conformal covariance studied here and the primarity addressed in ordinary CFT.

\se{The model: Free conformal invariant scalar field.}
In this section we present a model 
that shows the previously presented abstract structure. 
We shall consider the 
massless conformally coupled scalar field theory.
Here and in the next sections we shall consider only the four dimensional case, that's because many of the presented results hold only in that case.
Later on, we shall briefly discuss the difficulties that arise in generalizing the outcomes to other dimensions. 

Just to fix some notation let us remind that the classical equation of motion of the conformal Klein Gordon scalar field $\varphi$ 
on a spacetime $(M,\bg)$ is
\beq\label{eqmoto}
P_\bg = -\Box_\bg +\frac{1}{6} R_\bg, \qquad   P_\bg\varphi =0, 
\eeq
where $\Box_\bg$ is the d'Alembert opeartor constructed out of the four dimensional metric $\bg$ and $R_\bg$ is the Ricci scalar of the metric $\bg$.
 We start our analysis with the study of the interplay 
 between conformal transformations, the fundamental solutions and the microlocal spectral condition \cite{Radzikowski, BFK}.

\ssb{Conformal transformation of the fundamental solutions}
Let us start recalling the transformation law satisfied by the operator 
$P_\bg$
under conformal embeddings.

\lemma{%
Let $\psi$ be a conformal embedding of $(\Mi,\bg_1)$ into $(\Mii,\bg_2)$, consider the corresponding weighted transformations $\psi^{(3)}_*$ and $\psi^{(1)}_*$ of test functions thought as mappings from $C^\infty_0(\Mi)\to C^\infty_0(\psi(\Mi))\subset C^\infty_0(\Mii)$. 
The following equivalence holds for every $f$ in $C^\infty_0(\Mi)$:
\beq\label{Ptransform}
P_{\bg_2}(\psi^{(1)}_*(f))= \psi^{(3)}_* (P_{\bg_1} (f)).
\eeq }
\proof{Because of the support properties of $f$ we know that the supports of the following smooth functions,
$\psi^{(1)}_*(f)$ and $\psi^{(3)}_* \circ P_{\bg_1} (f)$, are contained in $\psi(\Mi)$.
Hence we can restrict our attention to the image of $\Mi$ under $\psi$, namely to the spacetime $(\psi(\Mi),\bg_1)$.
Furthermore the conformal embedding $\psi$ becomes a 
standard conformal transformation 
if restricted to $\psi(\Mi)$, and the proof of that proposition descends straightforwardly by means of a  direct computation (a detailed analysis is contained in the appendix D of Wald's book \cite{WaldBook}).
}

We can relax the hypotheses written above and use as test functions only the smooth functions.
In this case the equivalence \nref{Ptransform} works if restricted to the image $\psi(\Mi)\subset \Mii$.
Another important extent of the transformation law of the wave operator $P_\bg$ we would like to stress is its interplay with weighted test functions.
Actually, because of the presence of the conformal factor in the transformation law of the operator defining the equations of motion we have that $P_\bg$ maps test functions of weight $1$ into test functions of weight $3$.

\vsp
In a globally hyperbolic spacetime $(M,\bg)$,
the advanced / retarded fundamental solutions $\De_\pm$ of the partial differential equation $P_\bg \phi =0$ are the unique maps from $C_0^\infty(M)$ to $C^\infty(M)$ such that $P_\bg \De_\pm f=f$ and the domains of $\De_\pm f$ are contained in the causal future / past of the support of $f$ respectively  $\supp \De_\pm(f)\subset J^\pm(\supp f)$. For the issues regarding the uniqueness see \cite{BAR}. 

Let us study the transformation law enjoyed by the fundamental solutions under conformal embeddings and hence by the causal propagator.
\lemma{Let $\psi$ be a morphism in $\loc$, hence $\psi$ is a conformal embedding between $\psi:(M,\bg)\to(M',\bg')$, let $\De_\pm$ and $\De_\pm'$ be the uniquely defined advanced/retarded fundamental solutions of $P_\bg$ and $P_{\bg'}$. Consider the following operators from $C^\infty_0(\psi(M))$ to $C^\infty(\psi(M))$: 
$$
 \De_\pm^\psi := {\psi}_*^{(1)} \circ \De_\pm\circ {{\psi}_*^{(3)}}^{-1}
$$
then $\De_\pm^{\psi}$ are the uniquely defined advanced / retarded fundamental solutions of $P_{\bg'}$ in $(\psi(M),\bg')$.
Furthermore $\De_\pm^{\psi}=\chi(\psi(M)){\De'_\pm}_{|C^\infty_0(\psi(M))}$, where $\chi(\psi(M))$ is the characteristic function of $\psi(M)$.}
\proof{$(\psi(M),\bg')$ is a global hyperbolic subspace of $(M',\bg')$, then, in order to show that 
$\De^\psi_\pm$ are the advanced / retarded fundamental solutions of $P_{\bg'}$ in $(\psi(M),\bg')$,
we have to check two properties, the first one is that $P_{\bg'} \De^\psi_\pm f=f$ and the other one is that the support of $\De^\psi(f)\subset J^\pm(\supp f)|_{\psi(M)}$ for every $f$ in $C^{\infty}_0(\psi(M))$.
First of all, consider the following chain of equalities valid in $\psi(M)$ for every $f'\in C^\infty_0(\psi(M))$ and $f={\psi^{(3)}_*}^{-1}(f')$:
$$
f'={\psi^{(3)}_*}(f)=\psi^{(3)}_* \circ P_{\bg} (\De_\pm f)=
P_{\bg'}\circ \psi^{(1)}_*(\De_\pm f)= 
P_{\bg'}\at \De_\pm^\psi \circ {\psi}_*^{(3)}(f)\ct=
P_{\bg'}\at \De_\pm^\psi (f')\ct.
$$ 
The second step is to check that the domain property are preserved by $\psi$. Nonetheless the properties of $\psi$ assure the validity of the following chain of inclusions,
$$
\supp \De^\psi_\pm f' = \psi(\supp \De_\pm f) \subset
\psi(J^{\pm}(\supp f))\subset J^{\pm}(\psi(\supp f))
$$ in $\psi(M)$. Furthermore, $\psi$ maps causal curves into causal curves preserving the orientation and from this it descends the last inclusion.
}

The causal propagator $E$ is defined as the advanced minus retarded fundamental solution $E=\De_+-\De_-$, it is a distribution on compactly supported smooth functions uniquely defined in a globally hyperbolic spacetime once $P_\bg$ is given. 
It can be seen as map form $C^\infty_0(M)$ to $C^\infty(M)$ namely the set of solutions of $P_\bg \phi=0$.

\noindent
Knowing the interplay between advanced, retarded fundamental solutions and conformal embeddings, we can derive straightforwardly the way in which the causal propagator $E$  transforms under conformal transformation, i.e. 
\lemma{%
\label{propagator}
Let $\psi$ be a morphism in $\loc$ between the two elements $(M,\bg)$, $(M',\bg')$ of $\loc$,
then $\chi(\psi(M))E'(\psi_*^{(3)}(f))=\psi_*^{(1)}(E(f))$ for any $f\in C_0^\infty(M)$.  }

\vsp 
\noindent
The two point functions of Hadamard type play a distinguished role in the formulation of a quantum field theory in curved spacetime \cite{KayWald}. 
From the work of Radzikowski \cite{Radzikowski} and Brunetti, Fredenhagen and K\"ohler \cite{BFK} 
we know that an Hadamard two-point function is characterized by the microlocal spectral condition. 
Hence we shall say that a two-point distribution $\om_2$ is of Hadamard type if its antisymmetric part corresponds to the causal propagator and if it satisfies the microlocal spectral condition, which means that the wave front set of $\om_2$ has a certain form:
\beq\label{microlocal}
\WF(\om_2)=\ag(x_1,k_1,x_2,k_2)\in T^*M\setminus\{0\} | (x_1,k_1)\sim(x_2,k_2), k_1\in V_+ \cg,
\eeq
where $(x_1,k_1)\sim(x_2,k_2)$ if it exists a null geodesics $\ga[0,a]\to M$ such that $\ga(0)=x_1$ and $\ga(a)=x_2$ and 
$k_1$ is the cotangent, coparallel vector to the geodesic at $x_1$ while $k_2$ is equal to the parallel transport along $\gamma$ of $-k_1$ on $x_2$. 
The next preliminary task we have to accomplish is to give the transformation rule for the Hadamard two-point function under conformal embeddings. While we have already seen that 
the causal propagator satisfies an homogeneous transformation rule we
would like to see what happens to the symmetric part of an $\om_2$ of Hadamard type.

\lemma{\label{trasfomega} Let $\psi$ be a morphism in $\loc$ from $(M,\bg)$ to $(M',\bg')$ and $\om_2$ a distribution on $C_0^\infty(M\times M)$ that satisfy the microlocal spectral condition then, consider
$$
\om_2^\psi(f,g) := \om_2({\psi^{(3)}_*}^{-1} f , {\psi^{(3)}_*}^{-1} g ).$$ 
$\om_2^\psi$  is  a  distribution on $C^{\infty}_0(\psi(M)^2)$ and it satisfy the microlocal spectral condition on $(\psi(M),\bg')$.
}

\proof{Since $\psi_*^{(3)}$ is a smooth invertible map from $C^{\infty}_0(M)$ to $C^{\infty}_0(\psi(M))$, $\om_2^\psi$ is a distribution.
Let us analyze its wave front set of $\om_2^\psi$ in $(\psi(M),\bg')$; 
the definition of wave front set does not depend on the metric $\bg'$, we have simply to analyze the relation between 
$M$ and $\psi(M)$.
Since the $\psi_*^{(3)}$ is smooth and invertible, and since  $\psi$ is a diffeomorphisms we can immediately conclude that 
$(x_1,k_1,x_2,k_2)$ is an element of $\WF(\om_2^\psi)$ if and only if $(\psi^{-1}(x_1),\psi_*^{-1}(k_1),\psi^{-1}(x_2),\psi_*^{-1}(k_2))\in (\WF(\om_2))$. Here $\psi_*^{-1}:T_{\psi{(p)}}{\psi{(M)}^* \to T_pM^*}$ defined in the standard way. 
We have to show that $(x_1,k_1)\sim(x_2,k_2)$ in $(\psi(M),\bg')$.
To this end we are seeking for a future directed null geodesic $\ga'$ in $\psi(M)$ whose extreme points are $x_1$ and $x_2$ and whose cotangent vector in $x_1$ is $k_1$ and in $x_2$ is $-k_2$.
Notice that, having $(\psi^{-1}(x_1),\psi_*^{-1}(k_1))\sim(\psi^{-1}(x_2),\psi_*^{-1}(k_2))$ in $(M,\bg)$, 
it exists a future directed null geodesic $\ga$ with such properties in $(M,\bg)$. 
Because of the properties of the conformal embedding, $k_1$ and $k_2$ are also null vectors in $(\psi(M),\bg')$.
Since $\psi$ is an orientation and time orientation preserving conformal embedding, $\ga'=\psi(\ga)$ turns out to be also  a future null geodesics in $\psi(M)$, furthermore, 
let $\la$ and $\la'$ be the affine parameters of $\ga$ and of $\psi(\ga)$, 
then $\frac{d\la'}{d\la}=c\Om^2$ where $c$ is a constant and $\Om$ is the conformal factor of $\psi$.
Notice that if $\psi_*^{-1}k_1$ is a cotangent vector of $\ga$ in $\psi^{-1}(x_1)$, $k_1$ has to be the cotangent vector of $\psi(\ga)$ in $x_1$, the same also holds for $-k_2$ in $x_2$.
Finally, since the orientation is preserved by $\psi$, the thesis turns out to be proved.
}

\noindent
The singular structure of an Hadamard two point function, called Hadamard parametrix, is fixed \cite{KayWald}, to proceed with our analysis it will be useful to analyze it in more details.
The Hadamard parametrix $H$ has the following expansion in a small geodesically convex neighborhood containing the points $x$ and $y$:
\beq\label{parametrix}
H(x,y)=\frac{1}{8\pi^2}\at\frac{u(x,y)}{\si_\epsilon(x,y)}+v(x,y)\log\frac{\si_\epsilon(x,y)}{\mu^2}\ct
\eeq
where $u$ and $v$ are certain smooth functions that depend only on the geometry of the spacetime $(M,\bg)$, once the equations of motion are chosen and 
$\sigma_\epsilon=\sigma+i (T(x)-T(y)) \epsilon+\epsilon^2/2 $, where $T$ is any time function \cite{KayWald} and $\sigma$ is half of the  squared geodesical distance between $x$ and $y$, taken with sign. 
We shall give further details on the local construction of $u$ and $v$ in the appendix.
The Hadamard parametrix depends on the dimensional parameter $\mu$, 
we shall fix this parameter once and for every spacetime in $\loc$.
Finally we would like to analyze the difference of the singular structures in the sense of the following lemma.

\lemma{
Let $\psi$ be a morphism in $\loc$ between the two elements $(M,\bg)$, $(M',\bg')$.
Let $H$ and $H'$ be the Hadamard parametrix respectively on  two geodesically complete neighborhood $\cO$ of $M$ and $\cO'$ of $\psi(M)$ such that $\cO'\subset\psi(\cO)$
then
$$
H({\psi^{(3)}_*}^{-1} f , {\psi^{(3)}_*}^{-1} g )-H'(f,g) = \int_{{\cO'}\times \cO'} f(x) A(x,y) g(y)\;  d\mu_{\bg'}(x) d\mu_{\bg'}(y)
$$
where $A(x,y)$ is a smooth symmetric function on ${\cO'}\times \cO'$ and $f,g\in C^{\infty}_0({\cO'}\times \cO')$. Furthermore, in general it is non vanishing, 
and its coinciding point limit is 
$$
A(x,x)=\frac{1}{(12\pi)^{2}} \at R_{\bg}(\psi^{-1}(x)) - \Om_\psi^2(x) R_{\bg'}(x)\ct,
$$ 
where $\Om_\psi$ is the conformal factor associated to $\psi$.
}
\proof{%
The distribution $H$ satisfy the microlocal spectral condition and its antisymmetric part corresponds to the causal propagator hence, also because of the preceding lemma, 
$$H^\psi(f,g):=H({\psi^{(3)}_*}^{-1} f , {\psi^{(3)}_*}^{-1} g ) $$ is of Hadamard type in $(\psi(M),\bg)$ too. 
From this property it is clear that $H^\psi-H'$ must be a smooth function. 
In the equation \nref{limite} of the appendix we have shows that 
$A(x,x)$, has precisely the given form,
hence, since $A(x,y)$ is a smooth function it cannot vanish in general.   
Finally, because of the lemma \ref{propagator}, the causal propagator in $(M,\bg)$ is mapped to the causal propagator in $(\psi(M),\bg)$.
Since the antisymmetric part of $H$ correspond s to the causal propagator, it descends that the antisymmetric part of $A$ must vanish.
}
We would like to remark that $A(x,x)$ does not depend upon the dimensional parameter $\mu$ present in the short distance expansion of the Hadamard parametrix 
\nref{parametrix}.
%
%
Moreover, a change of the length
scale  $\mu$, does not affect the coinciding point limit of the $v$ coefficient.

\proposizione{\label{mu}
Consider a normal neighborhood $\cO$ and two four dimensional Hadamard parametrix $H$ and $H'$ defined on $\cO$, that differs by the length scale $\mu$ and $\mu'$ then
$$
\lim_{x\to y} H(x,y)- H'(x,y) = 0.
$$
}
\proof{
The difference $H(x,y)-H'(x,y)$ is a smooth function and it is proportional to $(\log \mu-\log \mu')\;v(x,y)$, hence the proposition descends from the analysis of the coinciding point limit of the $v$ coefficient performed in the appendix, where it is shown that $v(x,x)$ vanishes.}
This result does not hold in general considering the coinciding point limit of the derivatives of fields or in dimensions different then four as we shall briefly discuss later.
We would like to stress that this is an important issue for having conformally covariant Wick powers.

\ssb{Quantization as a functor}
In \cite{BFV} it was shown that the quantization in terms of 
$C^*$ algebras $\gA(M,\bg)$ generated by the Weyl operators of the Klein Gordon field correspond to a functor $\gA$ from the category of isometrically related manifolds $\lloc$ to the category $\alg$.
We would like to briefly show that in the case of massless conformally coupled Klein Gordon fields the functor $\gA$ can be extended as a functor between $\loc$ and $\alg$ as described in the section \ref{functor}.
The difference between what we are considering here and the previously given picture \cite{BFV} is that in the definition of $\loc$, we have admitted conformal embeddings as morphisms between the elements of $\lloc$ too.
Hence we have simply to check the covariance of $\gA$ with respect to the larger group of morphisms of $\loc$.
In the sense of the discussion presented in section \ref{functor} we have to show that, being $\psi:(M,\bg)\to (M',\bg')$
 a conformal embedding in $\loc$,  
there exists a corresponding morphism 
$\alpha_\psi:\gA(M,\bg)\to\gA(M',\bg')$ such that 
$\gA(\psi(M,\bg))=\alpha_\psi(\gA(M,\bg))$. 

We shall skip many details that can be easily reconstructed knowing the results of \cite{dimock,BFV}.
For our purpose it will be sufficient to know that the morphism $\alpha_\psi$ can be straightforwardly constructed once 
a symplectic map between the two symplectic  spaces $(\cS(M,\bg),\si)$ and $(\cS(M',\bg'),\si')$ is given.
To be more precise let us analyze the construction of $(\cS(M,\bg),\si)$.
Using the causal propagator and the differential operator defined above we can construct the set of wave functions $\cS$ as follows:
$$
\cS(M,\bg) := E(C^\infty_0(M)).
$$
$\cS(M,\bg)$ can be equipped with a symplectic form defined in the following way.
Let $\varphi_f=Ef$ then, since the spacetime $(M,\bg)$ is globally hyperbolic, consider the following non degenerate symplectic form
$$
\sigma(\varphi_f,\varphi_g)=\int_\Si \at \varphi_f \pa_a \varphi_g - \varphi_g \pa_a  \varphi_f \ct n^a d\mu_\Si
=\int f (Eg)  d\mu_\bg
$$
where $\Si$ is a Cauchy surface, moreover $\sigma$
 is independent on the particularly chosen Cauchy surface $\Si$.
Furthermore $n$ is the unit vector normal to $\Si$, $\mu_\bg$ is the volume element induced by the metric $\bg$, and $\mu_\Si$ is the volume element restricted to the hypersurface $\Si$.

We already know that for every isometric embedding $\psi_0:(M,\bg)\to(M',\bg')$ it exists  a symplectic map from $(\cS(M,\bg),\sigma)$ to $(\cS(M',\bg'),\sigma')$.
A similar symplectic map exists also for a conformal embedding
$\psi:(M,\bg)\to(M',\bg')$.
In fact, from the transformation properties of the causal propagator seen in the lemma \ref{propagator}, we have that for every 
 $\varphi_1$ and $\varphi_2$ in $\cS(M,\bg)$  
$$
\sigma'(\psi^{(1)}_*(\varphi_1),\psi^{(1)}_*(\varphi_2))=\sigma(\varphi_1,\varphi_2).
$$
It is now a simple task to construct the automorphism
$\alpha_\psi$ from $\gA(M,\bg)$ to $\gA(M',\bg')$ in the same way as in \cite{BFV}.
Hence $\gA$ can be promoted as conformally covariant functor.

\se{Fields as natural transformations}
In order to build more interesting examples it is important to have an algebra of local observables  that encompasses more complicated objects as the powers of fields and the component of the stress tensor.
Here we shall remind the construction of the fields algebra  as presented in the book \cite{Wald} and then we would like to show that that scalar field is really a natural transformation between two functors.

\ssb{The CCR algebra} 
We would like to follow the algebraic approach so the starting point is the abstract $*-$algebra $\cA(M,\bg)$ generated by the identity $\bI$ and the  smeared quantum fields $\varphi(f)$, where $f$ is a test function (a smooth compactly supported function contained in the set denoted by $\cD(M)$).
Furthermore the abstract fields $\varphi(f)$ must satisfy the following further requirements  
\begin{itemize}
\item[(i)]	$\varphi(\alpha_1 f_1+\alpha_2 f_2)=\alpha_1\varphi(f_1)+\alpha_2 \varphi(f_2),$ where $\alpha_1,\alpha_2\in \bC$;
\item[(ii)] $\varphi(f)^* = \varphi(\overline{f}) $;
\item[(iii)] $\varphi(P_\bg f)=0$;
\item[(iv)] $\varphi(f_1)\varphi(f_2)-\varphi(f_2)\varphi(f_1)=iE(f_1,f_2) \bI$,
\end{itemize}
\noindent
where, $E$ is the causal propagator of the massless conformally coupled Klein Gordon field, whose equation of motion is given by the operator $P_\bg$ given in \nref{eqmoto}.
 The sets of $\cA(M,\bg)$ with the algebraic morphisms form a category $\talg$.
We would like to show that the abstract field $\varphi$ can be interpreted as a natural transformation between that category and $\testf^3$.

\proposizione{%
\label{funtore}
$\cA$ is a functor between the two categories $\testf^3$ and $\talg$, in fact: to every $(M,\bg)$ it is possible to associate $\cA(M,\bg)$ and be $\psi$ a conformal embedding between $(M,\bg)$ and $(M',\bg')$
$\cA(\psi)$ is defined as the morphism that acts on the fields in the following way
\beq\label{fieldmorphism}
\alpha_\psi(\varphi(f_1)\dots\varphi(f_n)):= \varphi'(\psi^{(3)}_*(f_1)\dots\psi^{(3)}_*(f_n))\;,
\eeq
where $\varphi$, $\varphi'$ are the fields that generate $\cA(M,\bg)$ and $\cA(M',\bg')$ respectively. 
}\\

The proof of the present proposition descends form the definitions given above, from the transformation rules of the causal propagator and from the composition rules of the morphisms between two algebras.
Moreover, exploiting the definition of $\cA$ and $\cD$ and using \nref{fieldmorphism} for one single field, we also have the following proposition

\proposizione{%
\label{funtore2}
The scalar field $\varphi$ is a natural transformation between the category $\testf^{3}$ and $\talg$ and hence it is a locally covariant  conformal field of weight $1$.
}\\

The difference in the weights between the field and the test functions can be understood exploiting the present heuristic representation of the field
$$
\varphi(f):=\int_M \varphi(x) f(x) d\mu_\bg,
$$
and considering the transformation rule enjoyed by the measure $\mu_\bg$ under conformal transformations. 

\ssb{Extension to the local algebra of fields and Wick monomials\label{extalg}}
As shown in \cite{DF,HW01}, in order to study the Wick monomials we have to extend the algebra $\cA(M,\bg)$ to a bigger one, that we shall indicate as $\cW(M,\bg)$. In this respect we follow the notation and construction introduced in \cite{HW01} referring  to that paper for technical details.
Essentially the normal ordered fields, when evaluated on states satisfying the microlocal spectral condition, turn out to be 
distributions with certain wavefront sets. 
We can then smear them with more singular objects, namely the compactly supported distributions characterized by a particular wave front set.
The normal ordering prescription plays a distinguished role in this construction, we would like to remind its definition.
The normal ordering with respect to the Hadamard singularity $H$ (where a unit of measure $\mu$ is chosen) is defined as follows
\beq{\label{Wn}}
:\varphi_n(x_1)\dots \varphi(x_n):_H:= \left.\frac{\de^n}{i^n\de f(x_1)\dots \de f(x_n)}  \; \exp \at \frac{1}{2} H(f\otimes f)+ i\varphi(f) \ct\right|_{f=0}\;.
\eeq
The algebra $\cA(M,\bg)$ can now be enlarged allowing the smearing by more singular object then smooth functions in $C_0^\infty(M^n)$.
In particular, let us consider the following set
$$
\cT^n(M):=\ag t\in\cD'(M), t\text{ symm. }, \text{supp}(t) \text{ is compact }, \WF(t) \cap \overline{ V_+ \cup V_-} = \emptyset\cg\; ,
$$
where $V_\pm$ are the forwards or backwards light cones in $T^*M$ whose tip $x$ is in $M$.
The requirement on the wave front set of the elements of $\cT^n(M)$ is introduced in such a way that fields smeared by 
the distribution $t\in\cT^n(M)$ can be unambiguously  tested on states satisfying the microlocal spectral condition. For a more complete analysis on the subject we refer to the papers \cite{BF2000, HW01}.
The algebra $\cW(M,\bg)$ can now by defined as the $*$-algebra generated by the elements defined as in \nref{Wn} smeared by $t\in\cT^n(M)$.

\remark It can be shown combining the results in \cite{BF2000,HW01} that the algebra constructed in that way is independent on the choice of the Hadamard two point function $H$. In other words, substituting $H$ in the definition of the normal ordering with another two point distribution with the same singular structure, gives a set of generators of an isomorphic algebra.
Part of this freedom is encoded in the choice of the unit length $\mu$. It is in any case possible to add a smooth symmetric function to $H$ without really changing the $*$-algebra $\cW(M,\bg)$.

We are now ready to study the Wick monomials that are defined as the normal ordered products of fields smeared by some special test distributions.
More precisely, suppose to have a smooth function with compact support $C^\infty_0(M)$ then a Wick monomial $\varphi^n(f)$ of order $n$ can be defined as follows:
\beq\label{wickpowers}
:\varphi^n:_H(f):=\int  :\varphi(x_1)\dots \varphi(x_n):_H t_f(x_1,\dots,x_n) \;  d\mu_\bg(x_1) \dots d\mu_{\bg}(x_n)
\eeq
where  $t_f(x_1,\dots,x_n)$ is $f(x_1)\De(x_1,\dots,x_n)$ 
and $\De$ is the diagonal distribution $\De(x_1,\dots,x_n)$$=\de(x_1,x_2)$$\dots $$\de(x_{n-1},x_n)$.

The Wick powers defined in that way satisfy certain interesting properties, in particular they turn out to be locally covariant field in the sense of \cite{BFV}.
Another important extent showed by $:\varphi^k:_H$ is the almost homogeneous scaling under rigid dilations, where
 the non homogenous term is logarithmic in the scaling parameter.
Hollands and Wald have used an axiomatic approach, i.e., they have promoted  these and other physically  motivated properties 
to a set of axioms that every reasonable definition of Wick powers should satisfy.
In \cite{HW01}, they have furthermore shown that, the previously given definition for $\varphi^k$ is the unique one that 
satisfies the axioms up to the following renormalization freedom
\beq\label{redefinition}
\Tilde{\varphi}^k(x)=\varphi^k(x) + \sum_{i=1}^{k-2} C_{i}(x)  
\varphi^i(x)
\eeq
where $C_i(x)$ are classical fields depending on the parameter of the Lagrangian, and on the metric tensor, furthermore it is required that $C_i$ scale homogeneously under rigid dilation while the total field  $\varphi^k$ scales almost homogeneously, where the non homogeneous term must be of logarithmic type in the scaling parameter.
Hence, it is not possible to get rid of this non homogeneous logarithmic scaling behavior by a suitable choice of the renormalization constants $C_i(x)$.

\ssb{Wick monomials and conformal covariance}
It is known that the Wick monomials previously defined are locally covariant quantum fields in the sense of the analysis performed in \cite{BFV}. Here we would like to see that these fields are also locally conformal covariant.
Let's start our discussion analyzing the simplest case  of $\varphi^2(x)$. 
Here the freedom \nref{redefinition} consists of the following redefinition
\beq\label{phi2}
{\varphi}_\alpha^2(x)=:{\varphi}^2:_H(x) +\alpha R(x)
\eeq
where $R$ is the scalar curvature and $\alpha$ is a constant.

We would like to stress that this freedom is not included in the choice of a particular length scale $\mu$ in the Hadamard parametrix, in fact, as discussed in proposition \ref{mu},
the change of the length scale $\mu$ does not affect the expectation value of $:\varphi^2:_H$, while changing the parameter $\alpha$ in \nref{phi2} modifies its expectation value.

An interesting observation is the fact that both
$:{\varphi}^2:_H(x)$ and ${\varphi}_\alpha^2(x)$ scale homogeneously  under rigid dilations,
as can be seen from the transformation rules of the scalar curvature and the Hadamard singularity.
Let $H_\bg$ be the Hadamard singularity in the spacetime $(M,\bg)$, usually under rigid scaling $\lambda$ it should transform in the following way
$$
\la^{-2} H_{\la^{-2}\bg}(x,y)=H_{\bg}(x,y)+v_{\bg}(x,y) \log \la^2,
$$
notice that in the case under consideration $v_\bg(x,x)=0$, as can be seen form the appendix.
Furthermore, $R_\bg$ transforms homogeneously under rigid re-scaling too 
$$
\la^{-2}R_{\la^{-2}\bg}=R_\bg,
$$ 
hence the Wick monomial \nref{phi2} transforms homogeneously under rigid dilation.

\noindent
The second step in the analysis consists of testing $\varphi_\alpha^2$ under local transformation. 
Let $\psi$ be a conformal transformation from $(M,\bg)$ to $(M,\bg')$, then,  taking into account the transformation rule of the Hadamard singularity $H$ as given in the appendix, we have
$$
{\varphi'}_\alpha^2(\psi_*^{(2)}(f))=\varphi_\alpha^2( f) - \at  \frac{1}{(12\pi)^2}+\alpha\ct \int_M (R_\bg-(\Om\circ\psi)^2 R_{\psi\bg}) f  d\mu_\bg,
$$
where $\varphi_\alpha^2$  is the field on $(M,\bg)$ while 
${\varphi'}_\alpha^2$ is the one on $(M,\bg')$
The particular choice $\alpha=-1/(12\pi)^2$ makes the field conformally covariant.
We would like to see if this is the case also for more involved fields.
Namely we shall look for a particular redefinition of the Wick monomials, by a suitable choice of the renormalization constants $C_i(x)$ in \nref{redefinition}, to get rid of the non homogeneous behavior which is in general present in such cases.
We are going to show that this is the case by the following Theorem.

\teorema{Let $\varphi^k$ be a Wick power as given in \nref{wickpowers},
there is a non trivial choice of the renormalization constants $C_i$ in \nref{redefinition} that makes $\varphi^k$ a conformal locally covariant field with weight $k$ in the sense of the Definition \ref{conformalfield}.}

\proof{The proof is constructive: let us consider the following smooth function $B(x,y)= \frac{1}{2(12\pi)^2} (R_\bg(x)+R_\bg(y))$, then redefine the Wick monomials in the following way, 
$$
{\varphi}^k:=\;:\varphi^k:_{H+B}
$$
where
$$ 
:\varphi(x_1)\dots \varphi(x_k):_{H+B}= \left.\frac{\de^k}{i^k\de f(x_1)\dots \de f(x_k)}  \; \exp \at \frac{1}{2} (H+B)(f\otimes f)+ i\varphi(f) \ct\right|_{f=0}.
$$ 
The algebra generated using this new normal ordering is isomorphic to $\cW(M,\bg)$, the proof is similar to the one of the independence of the state given in \cite{HW01}; furthermore, it can be shown that $:\varphi:_{H+B}$ is related to $:\varphi:_H$ by a choice of the renormalization constants as in \nref{redefinition}.
The difficult part is to show that the Wick monomials defined with 
respect to the new normal ordering, satisfy the covariance condition with respect to the 
conformal embedding $\psi:(M,\bg)\to (M',\bg')$ in $\loc$ and its corresponding algebraic morphism $\alpha_\psi$ defined as in \nref{fieldmorphism}
$$
\alpha_\psi(:\varphi^k :_{H+B}(f))-:{\varphi'}^k :_{H'+B'}(\psi_{*}^{(4-k)}(f))=0.
$$
To this end, consider a general element $W$ of the Wick expansion of $:\varphi^k:_{H+B}(f)$
\begin{gather}\notag
W(x_1,\dots,x_k):=\int  \varphi(x_1)\dots\varphi(x_n) (H+B)(x_{n+1},x_{n+2}) \dots (H+B)(x_{k-1},x_{k})\\
t_f(x_1,\dots ,x_k)  
d\mu^1_\bg \dots d\mu^k_\bg
\label{element}
\end{gather}
where $t_f(x_1,\dots,x_k)= f(x_1)\De(x_1,\dots,x_k)$.
We would like to show that on $\psi(M)^k$
$$
S(f'):=\alpha_\psi(W(t_f)) - W'(t'_{f'})=0
$$
where  $W$  is as in \nref{element} and $W'(x_1,\dots,x_k )$ is the corresponding term of the expansion of ${:{\varphi'}^k:_{H'+B'}}(f')$ on $(\psi(M),\bg)$ with $f':=\psi_{*}^{(4-k)}(f)$.
First of all notice that $\alpha_\psi$ has no action on $(H+B)$ while 
$$
\alpha_\psi(\varphi(x))= \Om^{-1}(\psi(x))\varphi'(\psi(x)) \; .
$$
Hence 
\begin{gather}
S(f'):= \notag
\int  \varphi'(x_1)\dots\varphi'(x_n)  \\
\aq 
\Om^{-1}(x_{n+1})\dots \Om^{-1}(x_k)
(H+B)(x_{n+1},x_{n+2}) \dots (H+B)(x_{k-1},x_{k}) - \right.  \notag \\ \left.
(H'+B')(x_{n+1},x_{n+2}) \dots (H'+B')(x_{k-1},x_{k}) \cq \notag \\
f' \De'(x_1,\dots,x_k )
d\mu^1_{\bg'} \dots d\mu^k_{\bg'}  \notag
\end{gather}
where we have used the fact that $f(x_1)\De(x_1,x_2)=f(x_2)\De(x_1,x_2)$.
The proof can be concluded using the analysis presented in the appendix \nref{limite}, hence
for $y$ in a geodesically convex neighborhood $O$ of the point $x$ in $\psi(M)$, we have that
$$
\lim_{y\to x}   \frac{1}{\Om(x)}(H+B)(\psi^{-1}(x),\psi^{-1}(y)) \frac{1}{\Om(y)} - (H'+B')(x,y) = 0\; .
$$
With this observation, the proof can be concluded.
}
The function $B$ does not depend on the length scale $\mu$ present in the Hadamard parametrix. Hence even if the regularization procedure depends on that length scale, it does not appear explicitly in
the Wick powers.
Once again, this is an unexpected result that permits the construction of an infinite series of conformally covariant fields in the four dimensional case.
On the other hand, considering a general Wick monomial that contains also derivatives the length scale $\mu$ becomes important, in the sense that it affects the expectation value of such monomial.

\ssb{Extension of the results on different dimensions}
In this subsection we would like to emphasize the difficulties that appear in a possible generalization of the found 
results to spacetimes with general dimension $d$ different than four.
We shall discuss some aspects of the two dimensional case and we shall stress the differences with the four dimensional case in particular.
We start recalling that, in analogy with \nref{eqmoto}, in a $d$ dimensional spacetime $(M_d,\bg_d)$ the conformal invariant fundamental scalar field $\varphi$ has to satisfy the following equation
$$
-\Box \varphi_d +\frac{d-2}{4(d-1)} R\;  \varphi_d =0
$$
where $\Box$ is the d'Alembert operator and $R$ the scalar curvature of $(M_d,\bg_d)$.

Following the discussion presented above for the four dimensional case and the propositions  \ref{funtore} and \ref{funtore2} in particular, it is a straightforward task to construct the CCR algebra of this field and to interpret it as a functor.
Similarly, the conformal covariance of the microlocal spectral condition on $d$ dimensions can be shown to hold along the guidelines given in Lemma \ref{trasfomega}.
The difficulties arise in considering the extended algebra of fields
$\cW_d$ as done in four dimensions in the subsection \ref{extalg}.
This become manifest in the analysis of 
the transformation rules enjoyed by the
Wick powers $\varphi_d^k$ under conformal embeddings.
In order to touch this fact and to enlighten the difference it is 
helpful to consider once again the particular field $\varphi_d^2$, and the Hadamard parametrix $H_d(x,y)$ in particular.
In the even $d$ dimensional case, similarly to \nref{parametrix}, 
 the Hadamard parametrix takes the form
$$
H_d(x,y)= C_d\at \frac{u_d(x,y)}{\sigma_\epsilon ^{d/2-1}} + v_d(x,y) \log \frac{\si_\epsilon}{\mu^2}\ct,
$$
where $u_d$ and $v_d$ are again  smooth functions and $\sigma_\epsilon$ is half of the squared geodesic distance taken with sign regularized  as in \nref{parametrix}, $C_d$ is a dimensional dependent constant. 
For a detailed analysis of the Hadamard parametrix we refer to the paper \cite{Moretti03} and to the references therein. 

Notice that in the even dimensional case the Hadamard parametrix contains a length scale in the logarithmic part, and 
this length scale breaks the conformal covariance already at the level of $\varphi^2$. 
To see this extent explicitly 
consider two Hadamard different parametrix $H_d$ and $H_d'$ constructed respectively with $\mu$ and $\mu'$,
the difference between the two is simply the smooth function
$$
2 C_d\; v_d(x,y) \log \frac{\mu}{\mu'},
$$
and the change in the expectation value of $:\varphi_d^2:_{H_d}$ is 
$2 C_d\;  v_d(x,x) \log \frac{\mu}{\mu'}. $

As already discussed above, in four dimensions, it happens that $v_4(x,x)=0$ and hence a change of $\mu$ has no effect on $:\varphi^2_4:_{H_d}$.
Unfortunately this is not a general fact, and usually $v_d(x,x)\neq 0$. This computation is particularly easy in two and six dimensions.
Being $v_d(x,x)\neq 0$, it  happens that the field $\varphi^2_d$ transforms non-homogeneously under rigid dilations where in the non-homogenous part a logarithmic term in the scaling parameter $\lambda$ appears.
Following the discussion of Hollands and Wald, it is then not possible to cancel this logarithmic term by a judicious choice of other renormalization constants. The same extent is shown by the others Wick powers $\varphi^k_d$.

On the other hand, in two dimensional conformal field theories , it is known that 
the fields $\varphi^k$ are only quasi-primary but not primary, and hence they cannot be thought as natural transformations in the sense discussed in the present paper.
%
%
%
%
%
As a final comment we would like to stress that the study of 
conformal covariance in general dimensions requires a detailed case by case analysis of the Hadamard coefficient $v_d$ that is out of the scope of the present paper.


\se{Final comments}

We have generalized the notion of generally covariant fields to  encompass the conformally covariant transformations. 
This was done exploiting the theory of category in a similar way as in \cite{BFV}. 
We have furthermore analyzed the case of the conformally coupled massless Klein Gordon field, studying its Wick powers. 
Particularly we have shown that, using in a suitable way the renormalization freedom, it is possible to get rid of the non homogeneous part carried by the conformal transformation of those fields.
In a certain sense the larger group of covariance reduces the renormalization freedom.
The situation presented here is different than the one given in 
\cite{BFV}, due to the presence of the weights in front of the fields.
It is indeed not possible to linearly combine fields with different weights without breaking the conformal covariance, unless position dependent coupling constants  are taken into account.

Before concluding the discussion we would like to give some simple examples of other type of fields that fit into the presented framework. 
As an example of conformally covariant field with non constant couplings consider 
$$
\lambda_1 :\varphi^4:_{H+B} + {(W^2)^{1/2} \lambda_2 } :\varphi^2:_{H+B} +  W^2\; \lambda_3
$$
where $\la_1,\la_2,\la_3$ are constants and $W^2$ is the square of the Weyl tensor ${W_{abc}}^d$ , namely $W^2={W_{abc}}^d {W^{abc}}_d$.
Such a field is a conformally covariant field in the sense of definition \ref{conformalfield} and its weight is $4$.


Other interesting cases arise taking into account fields  containing covariant derivatives. Usually that kind of fields are more complicated and it is difficult to draw some general conclusions because of the presence of quantum anomalies, but also because of the non homogeneous transformation rule enjoyed by the covariant derivatives.
Nevertheless, also in that case it is possible to construct fields that 
are conformally covariant, provided a renormalization constant is chosen. As an example of these fields consider
$$
-:\nabla_a\varphi\Box\varphi:_H+\frac{R_\bg}{12} \nabla_a:\varphi^2:_H,
$$
notice that their classical counterparts are quite trivial since they vanish. 
On the other hand, also in that case there is a renormalization freedom of the form \nref{redefinition};
 we can add to it an homogeneous scaling constant $C$.
If $C$ is chosen as $C(x)=-2 \nabla_a v_1(x,x)$ \footnote{For technical details we refer to \cite{Moretti03,HW05}} 
that field turns out to vanish also quantum mechanically and, even if it is a trivial field, it can be interpreted as a conformally covariant field in the sense of definition \ref{conformalfield}.

\section*{Acknowledgements.} 
I would like to thank Romeo Brunetti, Claudio Dappiaggi,  Klaus Fredenhagen, Valter Moretti and Karl-Henning Rehren for useful discussions, suggestions and comments on the topic.
This work has been supported by the German DFG Research Program SFB 676.

\appendix
\se{Some technical computations}
\ssb{Transport equations}
The coefficients $u$ and $v$ given in the Hadamard parametrix \nref{parametrix} are symmetric smooth functions \cite{Moretti} that 
satisfy the following relations:
$$
2\nabla\si(x,y)\nabla u(x,y) + (\Box_x\si-4) u(x,y)=0, \qquad -P_x v=0.
$$
Moreover the coefficient $u$ is twice the square root of the van Vleck Morette determinant $u=2\De^{1/2}$, for definition and details see \cite{DB,Friedlander,Fulling,Tadaki}.
Furthermore, on a geodesically complete neighborhood, the function $v$ can be expanded as follows
$$
v=\sum_{n=0}^p v_n\si^n + O(\si^n).
$$
We have truncated the series at some order $p$ because, in general, the whole series does not converge, unless the coefficients of the metric are analytic functions.
Furthermore, the coefficients  $v_n$ can be found, using the following two recursive relations valid for $n>0$
$$
2\bg(x) \nabla_x\sigma \nabla_x v_0 +
(\Box_x \si(x,y)-2) v_{0}  = P^{(x)}_\bg u(x,y)\;,
$$
$$
 2n\; \bg(x) \nabla_x\sigma \nabla_x v_n +
n\; (\Box_x \si(x,y)+2n-2) v_{n}  =P^{(x)}_\bg v_{n-1}(x,y)\;.
$$

\ssb{Transformation laws for the Hadamard coefficients}
Consider a conformal transformation $\psi:(M,\bg)\to(M,\bg')$ with conformal factor $\Om$.  Let $H$ and $H'$ be the Hadamard singularities, as given in \nref{parametrix}, on a $(M,\bg)$ and $(M,\bg')$ respectively.
For $y$ in a geodesically complete neighborhood of the point $x$,  we would like to compute the coinciding point limit of the subtraction 
$$
\frac{1}{\Om(x)}H(x,y)\frac{1}{\Om(y)}-H'(x,y).
$$
Because of the Lemma \ref{trasfomega} we know that the subtraction is a smooth function, hence we can compute the following limit directly
\beq\label{limite}
\lim_{y\to x} \frac{u(x,y)}{\Om(x)\sigma(x,y)\Om(y)} + \frac{v(x,y)}{\Om(x)\Om(y)} \log \sigma -\frac{u'}{\sigma'} - v' \log \si'= \frac{R_\bg(x)}{18\, \Om^2(x)}-\frac{R_{\bg'}(x)}{18},
\eeq
where we have used the following expansions around $x$. 
Let $\sigma^\mu=\nabla_x^\mu\sigma$, and $L_\mu :=\nabla_\mu \log \Om$ then we can write the Taylor expansion 
$$
\Om(y)=\Om(x)\at 1 -L_\mu\si^{\nu} + \frac{1}{2}\at L_{\mu\nu}+L_\mu L_\nu \ct \si^\mu\si^\nu\ct + O(\si^{3/2}).
$$
Furthermore using the notation of the book of Fulling \cite{Fulling}
$$
\sigma'(x,y)=\Om^2(x)\sigma(x,y)\at 1-L_\mu \si^\mu -\frac{1}{12}\at - 2\si L_\mu L^\mu +\at 8 L_\mu L_\nu + 4 L_{\mu\nu} \ct\si^\mu \si^\nu \ct\ct + O(\si^{5/2})
$$
and the short distance analysis of van Vleck Morette determinant \cite{DB} gives
\beq\label{vvm}
\De^{1/2}= 1-\frac{1}{12} R_{\mu\nu} \si^\mu\si^\nu + O(\si^{2}).
\eeq
Notice that, in the case under investigation, because of the expansion \nref{vvm}, and the recursive relations given before, $v_0(x,x)=v(x,x)=0$.
Plugging the expansions written above into the previous subtraction and knowing that $v(x,x)=0$, \nref{limite} holds.

\end{document}